\documentclass[%
 reprint,
 amsmath,amssymb,
 aps,
]{revtex4-1}

\usepackage{graphicx}
\usepackage{dcolumn}
\usepackage{bm}
\usepackage{float}
\usepackage{array}
\newcolumntype{M}[1]{>{\centering\arraybackslash}m{#1}}
\usepackage{diagbox}

\begin{document}

\preprint{APS/123-QED}

\title{Oscillate Boiling from Electrical Microheaters}

\author{Dang Minh Nguyen}
 \affiliation{School of Physical and Mathematical Sciences, Division of Physics and Applied Physics, Nanyang Technological University, Singapore}
\author{Liangxing Hu}
 \affiliation{School of Mechanical and Aerospace Engineering, Nanyang Technological University, Singapore}
\author{Jianmin Miao}
 \affiliation{School of Mechanical and Aerospace Engineering, Nanyang Technological University, Singapore}
\author{Claus-Dieter Ohl}
\affiliation{School of Physical and Mathematical Sciences, Division of Physics and Applied Physics, Nanyang Technological University, Singapore}
\affiliation{Otto-von-Guericke University, Institute for Physics, Magdeburg, Germany}

\date{\today}

\begin{abstract}
Oscillate boiling offers excellent heat transfer at temperatures above the Leidenfrost temperature. Here we realize an electrical microheater with an integrated thermal probe and resolve the thermal cycle during the high-frequency bubble oscillations. Thermal rates of $10^8\,$K/s were found indicating its applicability for compact and rapid heat transfer from microelectrical devices.
\end{abstract}

\pacs{Valid PACS appear here}

\maketitle


\paragraph*{Introduction}

The boiling crisis is one of the most challenging problems among thermal and fluidic instabilities \cite{lloveras2012boiling}. It takes place when the heat fed into a boiling heater overcomes a threshold and the nucleate boiling bubbles merge into a vapor film. At that moment the liquid loses the direct contact to the heater \cite{dhir1998boiling} because the vapor layer has a greatly reduced heat transfer coefficient. As a result, the heater's temperature rises, which may lead to a thermal induced breakdown of the heater. Research has extended the limits of the boiling crisis by micro-structuring the substrate \cite{liter2001pool,chen2009nanowires,chu2013hierarchically,rahman2014role,dhillon2015critical,choi2016large} or adding nanoparticles to the liquid (nanofluids) \cite{das2003pool,wen2005experimental,bang2005boiling,chopkar2008pool}.

Recently, Li {\em et al.}~\cite{li2017oscillate} reported a new boiling regime which might allow to overcome the boiling crisis through stable localized heating, i.e. the {\em oscillate boiling regime}. Their finding was that bubbles at the heater undergo stable high-frequency oscillations while remaining pinned. In particular, the bubble does not grow over time. This is achieved by shedding excess amount of gas and vapour by pinching off a microbubble during each oscillation. Additionally, a large scale flow is generated that cools the substrate thereby keeps the temperature distribution localized to the heater. The key feature to generate the oscillate boiling regime is a miniature and localized heat source. A typical size of the heater to operate in water is about $10\,\mu$m in diameter. Previously this was achieved by tightly focusing a continuous wave laser beam onto an absorbing substrate made of gold nanoparticles~\cite{li2017oscillate}. Yet, using laser light as a heat source carries two drawbacks: Firstly, due to optical loss from the focusing optics and through reflections at the gold layer, only about $20\,$\% of the laser power remains available for heating. Secondly, the optical setup to generate more than one bubble becomes rather complex and rules out practical scaling up of the phenomenon. In the present work we overcome these two restrictions (efficiency and scalability) by demonstrating that oscillate boiling can be achieved with an electric heater. The additional benefit is the ability to measure the instantaneous heater's temperature to elucidate the complex fluid mechanics in the novel regime. 

Previously, electric miniature heaters have attracted attentions, e.g \cite{chen2002coalescence,deng2003design,deng2006experimental,chen2006subcooled,xu2008effect,romera2008explosive}. They were used to create and study the dynamics of the conventional nucleate boiling, where the liquid is heated rapidly above the spinodal limit leading to an explosive vaporization. We are not aware of reports demonstrating a stably oscillating bubble characterizing the oscillate boiling regime.

\paragraph*{Heater fabrication}

\begin{figure}
\centering
\includegraphics[width=1.\linewidth]{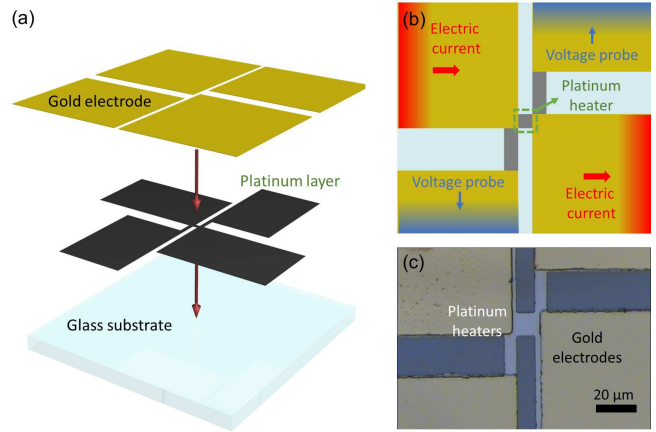}
\caption{Design of the electrical microheater. (a) The fabrication process. The platinum heating layer is deposited on top of glass substrate, followed by the gold electrode layer. (b) The four probe structure used to measure the heater's resistance. (c) Captured image of the microheater. The scale bar shows 20 $\mu$m}
\label{fig1}
\end{figure}

We have designed and fabricated a miniature heater to achieve the above mentioned temperatures on a sufficiently small scale. The heater consist of a $50\,$nm thick platinum (Pt) layer, deposited on a $200\,\mu$m thick glass substrate. The heater is contacted with a 600 nm thick gold layer deposited with magnetron sputtering (Fig. \ref{fig1}a). The gold electrodes have a high electrical conductance and large cross-sectional area to ensure a current supply with little resistive loss. In contrast, the high resistance from the thin 15 $\mu$m $\times$ 15 $\mu$m Pt heater creates a very localized heating -- approaching the conditions of the laser focus based heater~\cite{li2017oscillate}. Prior to the fabrication, the design was tested and optimized with finite element simulations \footnote{See supplementary materials Fig.S1 and Fig.S2 for detailed fabrication process and thermal simulation of the electrical heater}.

The temperature-dependent electrical resistance of the Pt-microheater allows to measure the heater's temperature during the oscillate boiling cycles. For this we have added two more electrodes through which we probe the resistance. For the temperature range we are interested in the resistance of Pt which increases linearly with temperature:

\begin{equation}
R = R_0 [1+\alpha (T-T_0)]\quad ,
\end{equation}

\noindent where $R_0$ and $R$ are the initial and current resistance; $T_0$ and $T$ are the initial and current temperature, respectively; $\alpha$ is the temperature coefficient of resistance, which is calibrated to 1.023 $\times$ 10$^{-3}$ K$^{-1}$ . The resistance $R$ to determine the temperature $T$ is obtained from two additional probes measuring the voltage drop across the heater as shown in  Fig. \ref{fig1}b.

\paragraph*{Experimental setup.}

In the experimental setup the substrate with the heater is the base of an acrylic container (10 mm $\times$ 10 mm $\times$ 12 mm). Two of the faces of this cuvette are replaced with microscope cover glass to ensure good imaging of the microscopic dynamics. The container is then filled with DI water to the height of approximately $7\,$mm.

The microheater is operated in a constant current mode using a current controller for laser diodes (Thorlabs LDC 220C). The instantaneous current is measured from the voltage drop across a 10$\,\Omega$ resistor connected in series with the heater. The heater's current and its resistance are recorded with a 12-bit oscilloscope (Lecroy HRO 64Zi).

The small size of the bubble and its rapid dynamics demand for a high-speed camera. We use a SA-X2 (Photron) which is operated between 300,000 and 400,000 frames/s using a 20x magnification lens (20x PTEM M Plan Apo, Edmund Optics). This gives to a resolution of $1\,\mu$m per pixel. Additionally, a hydrophone (ONDA, model HNR1000) is placed near the heater to capture the acoustic signal emitted from the bubble oscillation. The hydrophone is connected to a third channel of the oscilloscope.

\begin{figure}
\centering
\includegraphics[width=1.\linewidth]{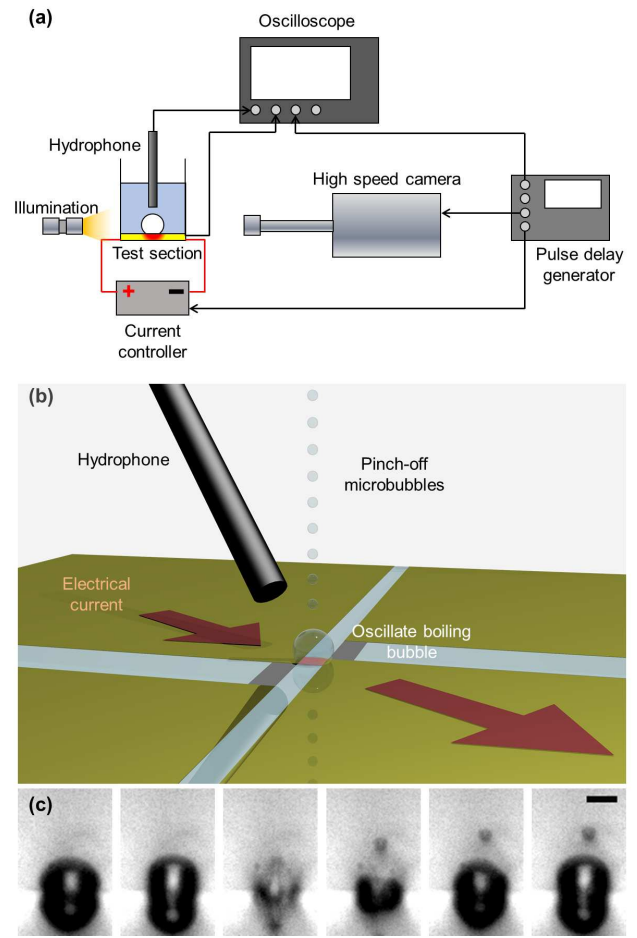}
\caption{(a) Sketch of the experimental setup to study the oscillate boiling heat transfer. (b) Zoomed-in image of the test section whereas the bubble is formed on top of the platinum heater (c) Consecutive images during one period of oscillation at a heating power of 90 mW. For better picture quality, a relatively large (yet unstably) oscillating bubble was chosen. Images of a stably oscillating boiling bubble are presented in Fig. S2 in the Supplementary Materials. The scale bar shows $20\,\mu$m and the frame rate is 400,000~fps.}
\label{fig2}
\end{figure}

Figure \ref{fig2}a illustrates the setup of the experiment. All electronic components (heater, power source, camera, oscilloscope, etc.) are connected to a common ground to reduce electrical noise. For the experiment the power source, camera and oscilloscope are simultaneously triggered with a digital pulse generator (BNC 575, Berkeley Nucleonics).

\paragraph*{Bubble oscillation.}

\begin{figure}
\centering
\includegraphics[width=0.9\linewidth]{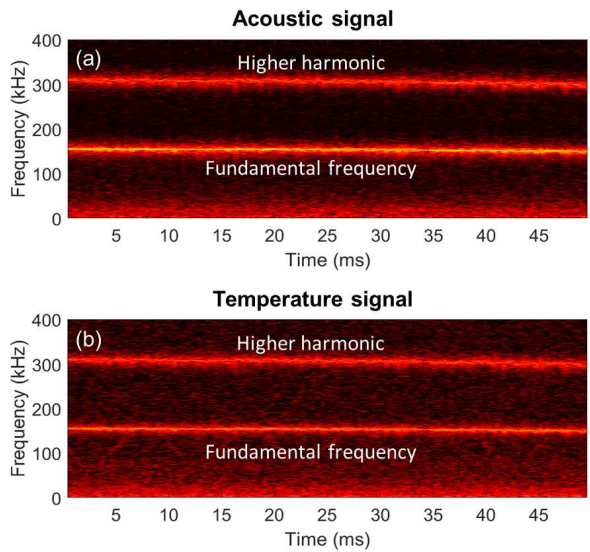}
\caption{(a) Spectrogram from a typical bubble's emitted acoustic signal recorded with a hydrophone. It reveals a strong fundamental frequency and higher harmonics. (b) The spectrogram of the heater's resistance is recorded simultaneously and shows a very similar power spectrum. The heating power used in this experiment is 86 mW.}
\label{fig3}
\end{figure}

After the power source is triggered, an initial vaporous explosion is observed, leaving behind a small bubble in contact with the microheater (Fig. \ref{fig2}b). Within a millisecond, this bubble enters an oscillation state at a fixed frequency. While oscillating a stream of microbubbles is ejected from the apex of the bubble. Figure \ref{fig2}c captures one oscillation period of a relatively large bubble. Here it is evident that a microbubble is pinched off during the bubble collapse. The main bubble then re-expands and a new cycle starts. The overall dynamics can also be obtained from the acoustic emission of the bubble. Figure~\ref{fig3}a depicts short-time power spectra of the acoustic emission recorded with hydrophone signal. The spectra is plotted as a function of time (spectrogram). The amplitude of the frequency components is indicated with the color; brighter colors depict larger amplitudes. For the hydrophone signal, Fig.~\ref{fig3}a, the fundamental oscillation frequency is about 160 kHz and constant over the covered measurement observation. Additionally, a second harmonic at 320 kHz is evident, which is well known for large amplitude bubble oscillations. In general, the fundamental frequency varies between 90 kHz and 300 kHz as a function of the bubble's size; it increases for smaller bubbles. 

\paragraph*{Temperature measurements.}

\begin{figure}
\centering
\includegraphics[width=1.\linewidth]{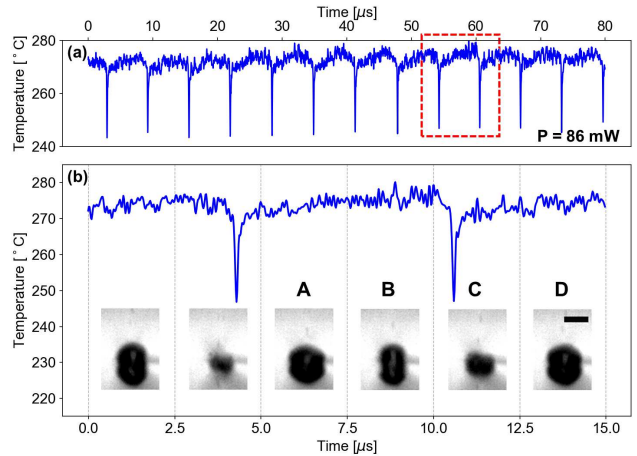}
\caption{(a) Heater's temperature during the oscillate boiling events using  an input power of 86 mW. (b) Zoomed-in temperature of the region marked with a dashed line in figure (a) and the corresponding bubble image. The scale bar is 20 $\mu$m.}
\label{fig4}
\end{figure}

The obvious advantages of an electrical system are its simpler control of the power input and the possibility to monitor the instantaneous heater temperature during the bubble oscillation. Figure \ref{fig4}a depicts such a recording. The signal consists of a series of cycles with a fast downward spike connected with a positive sloped signal of increasing temperatures in-between. Figure~\ref{fig4}b depicts a zoomed-in view of the thermal signal (see dashed line in Fig.~\ref{fig4}a) together with frames from a synchronous high-speed recording. The frames are recorded at the center of each interval indicated with the vertical lines, i.e. frame C showing a re-expanding bubble was taken at $t=11.25\,\mu$s. The drop of the heater temperature occurred shortly before this frame was recorded. The slope of the temperature change is remarkable: during the interval C in Fig.~\ref{fig4}b the heater temperature drops within approximately $50\,$ns by $30\,$K. Thus the heater experiences a cooling rate of $6\times 10^8\,$K/s! That is also the time when the microbubble is pinched off. The temperature recovery is slower, typically within  about $500\,$ns. After this fast recovery the temperature continues to rise at a constant slope, i.e. from $\approx 270^\circ\,$C to about $280^\circ\,$C before the next cycle starts.

Next we compare the thermal signal with the acoustic signal as shown in Fig. ~\ref{fig3}b. There the spectrogram of the thermal signal reveals close similarity with the acoustic signal. The larger noise floor of the thermal signal is due to the smallness of the electric signal and some noise pick-up.

In our previous work~\cite{li2017oscillate} it was argued that during the bubble collapse, e.g. between frame B and C of Fig.~\ref{fig4}b, a liquid jet impinges onto the substrate. Upon its contact with the hot surface, the heat transfer is strongly enhanced, leading to an instantaneous  vaporization of a minute amount of liquid and the subsequent vapor explosion is driving the re-expansion of the bubble. During this process the heater loses energy, resulting in a drop in temperature. Soon after, the heater recovers back to its initial temperature through Joule's heating and the cycle starts all over. 

This observation suggests that a sufficiently small electric heater is actively cooled by the oscillate boiling phenomenon. Applications which may profit from the cooling are integrated circuits or single high-power transistors operating above the Leidenfrost temperature. Here, oscillate boiling provides means to retain the device at a constant low temperature. It does so by utilizig the heat energy to drive to transport coold liquid to the device.

Comparing with the values from Baumeister {\em et al.} \cite{baumeister1973leidenfrost} for water as the fluid and platinum as the substrate, the Leidenfrost temperature is estimated to be $227 ^\circ\,$C. The heater's temperatures shown in Fig.\ref{fig4} are higher. Thus oscillate boiling remains operational above the Leidenfrost temperature which is in stark contrast to nucleate boiling. Thus oscillate boiling may offer a route to overcome the boiling crisis.

There have been reports where electrical heaters are used to study the thermal profile of a boiling bubble \cite{deng2006experimental,romera2008explosive}. Interestingly, Romera {\em et al.}~\cite{romera2008explosive} have also reported temperature oscillation from an electric microheater following the nucleate boiling event. They explained the bubble oscillations cycles with vaporization and condensation due to the thermal gradient in the surrounding subcooled liquid. We in contrast explain the cooling with liquid jetting; the bubble oscillation is essentially inertial cavitation (for a model see Ref.~\cite{li2017oscillate}). The most striking difference is that Romera {\em et al.} observed a few tens of thermal oscillations before they died out and the heater temperature rose. In contrast  we see oscillations lasting for millions of cycles. This difference can be explained that their heater is a thin strip which may allow the bubble to grow in the less confined direction destabilizing the oscillations.

\paragraph*{Parameter study.}

\begin{figure}
\centering
\includegraphics[width=1.\linewidth]{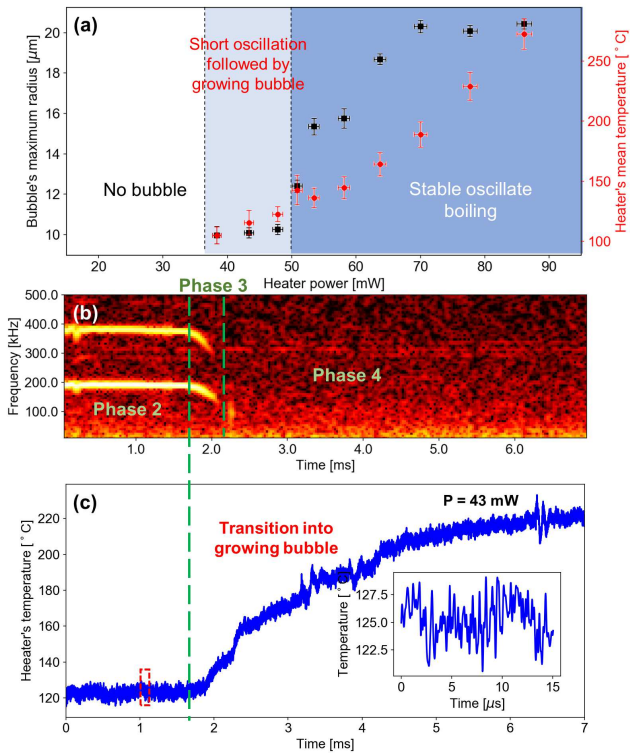}
\caption{(a) Properties of oscillate boiling bubble with different input power. In the unstable regime, the data is captured during the oscillations before the bubble grows thermally. Here the bubble's radius is defined as the radius of a hemispherical having the same volume (b) Spectrogram of the acoustic signal emitted from the bubble during the transition from oscillate boiling to growing mode. (c) Heater's temperature during the transition from oscillate boiling to growing mode. The inset is the enlargement of the red-dashed rectangle.}
\label{fig5}
\end{figure}

The key advantage of an electrical heater over the previous laser heat source is the ability to have a good control of the heating power by adjusting the electric current into the heater. Figure~\ref{fig5}a plots the maximum bubble radius and the heater's mean temperature of a square heater (15$\mu$ width) as a function of the electrical power. With increasing power, 3 distinct parameter regimes are identified: for electrical powers below 35 mW the vapor bubble explosion is absent and convective heating prevails. Once the electrical power is above 35 mW, a vapor explosion occurs followed by a small bubble undergoing volume oscillation (phase 1). However, unlike the one shown in Fig. \ref{fig4}, this oscillation shows neither the periodic microbubble pinch-off nor the temperature spikes from the liquid jet impact (phase 2). It lasts for several milliseconds and then decays as the bubble's size increases (phase 3). The oscillation eventually stops and the bubble continues to grow due to heating (phase 4) ~\footnote{See supplementary materials Video 2 for the recordings of this process}. During this transition, the heater's temperature increases from about 120$^\circ$C to more than 200$^\circ\,$C. Figure~\ref{fig5}b-c depicts the different phases. We call this regime unstable oscillate boiling to distinguish it from the stable oscillate boiling (shown in Fig. \ref{fig3}) where the oscillation has a constant frequency and lasts for hundreds of milliseconds. Likely this unstable behavior is the result that the  bubble accumulates residual gas from the liquid, i.e. through rectified diffusion and from residual gas in the vaporized water. We speculate that the smaller bubble in this regime does not collapse as violently. Then the jet is not reaching the hot substrate which stabilizes the bubble shape during re-expansion and avoids the microbubble pinch-off. This hypothesis is supported by the absence of thermal spikes in the heater's temperatures, see inset in Fig~\ref{fig5}c. The author believes that the observation of Romera {\em et al.}~\cite{romera2008explosive} falls into this unstable regime. 

Bubbles in the unstable oscillate boiling regime have 4 phases in common: The start through an explosive nucleation (phase 1) followed by an initial oscillation (phase 2). The frequency may or may not be constant. Phase 2 is followed by a decrease in oscillation amplitude and frequency while the bubble increases in size (phase 3). The oscillation then cease and the bubble grows thermally (phase 4). The duration of each phase varies greatly and were not able to attribute rerproducible trends. The Supplementary Materials demonstrate another representative example of an unstable oscillate boiling bubble which shares similar qualitative features but has largely different durations for the different phases \footnote{See Supplementary Materials section II.C for further discussion on the unstable nature of this regime.}.  

The regime of {\em stable oscillate boiling} can only be reached if the gas uptake of the bubble is balanced by the gas release through microbubble pinch-off. For the present geometry the electric power must be above $50\,$mW to achieve stable oscillate boiling. As the bubble enters this regime, its maximum effective radius increases, see Fig~\ref{fig5}a, leading to a stronger bubble collapse which favors liquid jet impact. The size of the bubble increases and the oscillation frequency decreases with increasing power. Table I compares the unstable oscillate boiling and stable oscillate boiling.

Qualitative and even quantitative similar oscillation frequencies in the stable regime are obtained with the laser as a heat source~\cite{li2017oscillate}. Yet, one needs to account that the two approaches lead to rather distinct heat distributions \footnote{See supplementary materials part II.D for detailed comparisons of two heating approaches.}. The laser spot forms a Gaussian-shaped heat distribution while the square-shaped electric resistor results in a more uniform heat distribution. The maximum power of the electrical design is $90\,$mW. Above this value the platinum thin film resistor is destroyed, likely due to melting. 

\begin{table}
\begin{center}
\begin{center}
\begin{tabular}  { | M{0.31\linewidth} || M{0.31\linewidth} | M{0.31\linewidth} | }
\hline
\textbf{Oscillate Boiling} & \textbf{Unstable} & \textbf{Stable}\\
\hline
\hline
Heating power range & 35mW - 50mW & 50mW - 90mW\\
\hline
Periodic microbubble pinch-off & no & yes \\
\hline
Rapid cooling during each cycle & no & yes\\
\hline
\hline
Phase 1 & \multicolumn{2}{M{0.62\linewidth} |}{Explosive nucleation results in a small oscillating bubble.} \\
\hline
Phase 2 & Either varying oscillation frequency or adjusts to a fixed frequency.& The bubble quickly adjusts to a fixed size and frequency. \\
\hline
Phase 3 &  Bubble increases in size while frequency drops.& N.A.\\
\hline
Phase 4 & Oscillations decays and bubble grows thermally. & N.A.\\

\hline
\end{tabular}
\end{center}
\label{Table1}
\caption{Distinguishing stable from unstable oscillate boiling regimes on a 15 $\mu$m $\times$ 15 $\mu$m size heater.}
\end{center}
\end{table}


Experiments were carried out with square heaters of different sizes, i.e. 10$\mu$m, 15$\mu$, and 20$\mu$m side length. The $10\,\mu$m and $15\,\mu$m-sized heaters show similar behaviors, namely with increasing electrical power a transition from convective heating to unstable and then to stable oscillate boiling. As expected the electrical power thresholds are lower for the 10$\mu$m-sized heater as compared to the 15$\mu$m heater~\footnote{See Supplementary Materials Fig. S7 for results with 10$\mu$m $\times$ 10$\mu$m heater.}. Interestingly the $20\,\mu$-sized heater can't produce stable oscillate boiling. Up to the point of electrical failure, only unstable oscillate boiling was found. This regime quickly transits into the thermal growth (phase 4). Again, this observation agrees with previously reported results with laser heating. There the focus had to be sufficient small to obtain stable oscillate boiling.

Figure~\ref{fig5}c depicts the heater temperature ($15\,\mu$) during the transition from periodic oscillation (phase 2) to thermal growth (phase 4). While the bubble oscillations decay and the thermal growth sets in the heater's temperature increases by about $100\,$K. We explain this temperature rise with the absence of advective cooling of the heater once the flow generated by the bubble oscillations ceases. The oscillate boiling regime provides means to sustain the heater at a lower operating temperature by advective cooling. 

In summary, the oscillate boiling regime can not only be achieved as previously shown by optical energy input but also with electrical heaters. First and foremost we demonstrated a suitable design to achieve this goal which also allowed to monitor the heater temperature. The experiments revealed during the bubble collapse enormous cooling rates of $6\times 10^8\,$K/s. We report on the heater's temperature for the three identified regimes of convective heating, unstable and stable oscillate boiling. At low electric power oscillate boiling is unstable. Thereby the average bubble volume increases over time due to accumulation of gas in the bubble. At higher power stable oscillate boiling is obtained. In this regime the heater's temperature is reduced considerably by the heat transfer due to jet impact of a liquid jet formed during the bubble collapse. Applications of the device may be large-scale heaters based on arrays of microheaters or the cooling of localized electric circuits through micro-jet impingement.

\begin{acknowledgments}

The research is supported by MOE Singapore (Tier 1 grant RG90/15). We thank Luong Trung Dung for his help with the experiments and Zeng Qingyun for helpful suggestions.

\end{acknowledgments}

\bibliography{bibliography}

\begin{thebibliography}{25}%
\makeatletter
\providecommand \@ifxundefined [1]{%
 \@ifx{#1\undefined}
}%
\providecommand \@ifnum [1]{%
 \ifnum #1\expandafter \@firstoftwo
 \else \expandafter \@secondoftwo
 \fi
}%
\providecommand \@ifx [1]{%
 \ifx #1\expandafter \@firstoftwo
 \else \expandafter \@secondoftwo
 \fi
}%
\providecommand \natexlab [1]{#1}%
\providecommand \enquote  [1]{``#1''}%
\providecommand \bibnamefont  [1]{#1}%
\providecommand \bibfnamefont [1]{#1}%
\providecommand \citenamefont [1]{#1}%
\providecommand \href@noop [0]{\@secondoftwo}%
\providecommand \href [0]{\begingroup \@sanitize@url \@href}%
\providecommand \@href[1]{\@@startlink{#1}\@@href}%
\providecommand \@@href[1]{\endgroup#1\@@endlink}%
\providecommand \@sanitize@url [0]{\catcode `\\12\catcode `\$12\catcode
  `\&12\catcode `\#12\catcode `\^12\catcode `\_12\catcode `\%12\relax}%
\providecommand \@@startlink[1]{}%
\providecommand \@@endlink[0]{}%
\providecommand \url  [0]{\begingroup\@sanitize@url \@url }%
\providecommand \@url [1]{\endgroup\@href {#1}{\urlprefix }}%
\providecommand \urlprefix  [0]{URL }%
\providecommand \Eprint [0]{\href }%
\providecommand \doibase [0]{http://dx.doi.org/}%
\providecommand \selectlanguage [0]{\@gobble}%
\providecommand \bibinfo  [0]{\@secondoftwo}%
\providecommand \bibfield  [0]{\@secondoftwo}%
\providecommand \translation [1]{[#1]}%
\providecommand \BibitemOpen [0]{}%
\providecommand \bibitemStop [0]{}%
\providecommand \bibitemNoStop [0]{.\EOS\space}%
\providecommand \EOS [0]{\spacefactor3000\relax}%
\providecommand \BibitemShut  [1]{\csname bibitem#1\endcsname}%
\let\auto@bib@innerbib\@empty
\bibitem [{\citenamefont {Lloveras}\ \emph {et~al.}(2012)\citenamefont
  {Lloveras}, \citenamefont {Salvat-Pujol}, \citenamefont {Truskinovsky},\ and\
  \citenamefont {Vives}}]{lloveras2012boiling}%
  \BibitemOpen
  \bibfield  {author} {\bibinfo {author} {\bibfnamefont {P.}~\bibnamefont
  {Lloveras}}, \bibinfo {author} {\bibfnamefont {F.}~\bibnamefont
  {Salvat-Pujol}}, \bibinfo {author} {\bibfnamefont {L.}~\bibnamefont
  {Truskinovsky}}, \ and\ \bibinfo {author} {\bibfnamefont {E.}~\bibnamefont
  {Vives}},\ }\href@noop {} {\bibfield  {journal} {\bibinfo  {journal}
  {Physical review letters}\ }\textbf {\bibinfo {volume} {108}},\ \bibinfo
  {pages} {215701} (\bibinfo {year} {2012})}\BibitemShut {NoStop}%
\bibitem [{\citenamefont {Dhir}(1998)}]{dhir1998boiling}%
  \BibitemOpen
  \bibfield  {author} {\bibinfo {author} {\bibfnamefont {V.}~\bibnamefont
  {Dhir}},\ }\href@noop {} {\bibfield  {journal} {\bibinfo  {journal} {Annual
  review of fluid mechanics}\ }\textbf {\bibinfo {volume} {30}},\ \bibinfo
  {pages} {365} (\bibinfo {year} {1998})}\BibitemShut {NoStop}%
\bibitem [{\citenamefont {Liter}\ and\ \citenamefont
  {Kaviany}(2001)}]{liter2001pool}%
  \BibitemOpen
  \bibfield  {author} {\bibinfo {author} {\bibfnamefont {S.~G.}\ \bibnamefont
  {Liter}}\ and\ \bibinfo {author} {\bibfnamefont {M.}~\bibnamefont
  {Kaviany}},\ }\href@noop {} {\bibfield  {journal} {\bibinfo  {journal}
  {International Journal of Heat and Mass Transfer}\ }\textbf {\bibinfo
  {volume} {44}},\ \bibinfo {pages} {4287} (\bibinfo {year}
  {2001})}\BibitemShut {NoStop}%
\bibitem [{\citenamefont {Chen}\ \emph {et~al.}(2009)\citenamefont {Chen},
  \citenamefont {Lu}, \citenamefont {Srinivasan}, \citenamefont {Wang},
  \citenamefont {Cho},\ and\ \citenamefont {Majumdar}}]{chen2009nanowires}%
  \BibitemOpen
  \bibfield  {author} {\bibinfo {author} {\bibfnamefont {R.}~\bibnamefont
  {Chen}}, \bibinfo {author} {\bibfnamefont {M.-C.}\ \bibnamefont {Lu}},
  \bibinfo {author} {\bibfnamefont {V.}~\bibnamefont {Srinivasan}}, \bibinfo
  {author} {\bibfnamefont {Z.}~\bibnamefont {Wang}}, \bibinfo {author}
  {\bibfnamefont {H.~H.}\ \bibnamefont {Cho}}, \ and\ \bibinfo {author}
  {\bibfnamefont {A.}~\bibnamefont {Majumdar}},\ }\href@noop {} {\bibfield
  {journal} {\bibinfo  {journal} {Nano letters}\ }\textbf {\bibinfo {volume}
  {9}},\ \bibinfo {pages} {548} (\bibinfo {year} {2009})}\BibitemShut {NoStop}%
\bibitem [{\citenamefont {Chu}\ \emph {et~al.}(2013)\citenamefont {Chu},
  \citenamefont {Soo~Joung}, \citenamefont {Enright}, \citenamefont {Buie},\
  and\ \citenamefont {Wang}}]{chu2013hierarchically}%
  \BibitemOpen
  \bibfield  {author} {\bibinfo {author} {\bibfnamefont {K.-H.}\ \bibnamefont
  {Chu}}, \bibinfo {author} {\bibfnamefont {Y.}~\bibnamefont {Soo~Joung}},
  \bibinfo {author} {\bibfnamefont {R.}~\bibnamefont {Enright}}, \bibinfo
  {author} {\bibfnamefont {C.~R.}\ \bibnamefont {Buie}}, \ and\ \bibinfo
  {author} {\bibfnamefont {E.~N.}\ \bibnamefont {Wang}},\ }\href@noop {}
  {\bibfield  {journal} {\bibinfo  {journal} {Applied Physics Letters}\
  }\textbf {\bibinfo {volume} {102}},\ \bibinfo {pages} {151602} (\bibinfo
  {year} {2013})}\BibitemShut {NoStop}%
\bibitem [{\citenamefont {Rahman}\ \emph {et~al.}(2014)\citenamefont {Rahman},
  \citenamefont {Ölçeroğlu},\ and\ \citenamefont
  {McCarthy}}]{rahman2014role}%
  \BibitemOpen
  \bibfield  {author} {\bibinfo {author} {\bibfnamefont {M.~M.}\ \bibnamefont
  {Rahman}}, \bibinfo {author} {\bibfnamefont {E.}~\bibnamefont
  {Ölçeroğlu}}, \ and\ \bibinfo {author} {\bibfnamefont {M.}~\bibnamefont
  {McCarthy}},\ }\href@noop {} {\bibfield  {journal} {\bibinfo  {journal}
  {Langmuir}\ }\textbf {\bibinfo {volume} {30}},\ \bibinfo {pages} {11225}
  (\bibinfo {year} {2014})}\BibitemShut {NoStop}%
\bibitem [{\citenamefont {Dhillon}\ \emph {et~al.}(2015)\citenamefont
  {Dhillon}, \citenamefont {Buongiorno},\ and\ \citenamefont
  {Varanasi}}]{dhillon2015critical}%
  \BibitemOpen
  \bibfield  {author} {\bibinfo {author} {\bibfnamefont {N.~S.}\ \bibnamefont
  {Dhillon}}, \bibinfo {author} {\bibfnamefont {J.}~\bibnamefont {Buongiorno}},
  \ and\ \bibinfo {author} {\bibfnamefont {K.~K.}\ \bibnamefont {Varanasi}},\
  }\href@noop {} {\bibfield  {journal} {\bibinfo  {journal} {Nature
  communications}\ }\textbf {\bibinfo {volume} {6}} (\bibinfo {year}
  {2015})}\BibitemShut {NoStop}%
\bibitem [{\citenamefont {Choi}\ \emph {et~al.}(2016)\citenamefont {Choi},
  \citenamefont {David}, \citenamefont {Gao}, \citenamefont {Chang},
  \citenamefont {Allen}, \citenamefont {Wang},\ and\ \citenamefont
  {Chang}}]{choi2016large}%
  \BibitemOpen
  \bibfield  {author} {\bibinfo {author} {\bibfnamefont {C.-H.}\ \bibnamefont
  {Choi}}, \bibinfo {author} {\bibfnamefont {M.}~\bibnamefont {David}},
  \bibinfo {author} {\bibfnamefont {Z.}~\bibnamefont {Gao}}, \bibinfo {author}
  {\bibfnamefont {A.}~\bibnamefont {Chang}}, \bibinfo {author} {\bibfnamefont
  {M.}~\bibnamefont {Allen}}, \bibinfo {author} {\bibfnamefont
  {H.}~\bibnamefont {Wang}}, \ and\ \bibinfo {author} {\bibfnamefont {C.-h.}\
  \bibnamefont {Chang}},\ }\href@noop {} {\bibfield  {journal} {\bibinfo
  {journal} {Scientific reports}\ }\textbf {\bibinfo {volume} {6}} (\bibinfo
  {year} {2016})}\BibitemShut {NoStop}%
\bibitem [{\citenamefont {Das}\ \emph {et~al.}(2003)\citenamefont {Das},
  \citenamefont {Putra},\ and\ \citenamefont {Roetzel}}]{das2003pool}%
  \BibitemOpen
  \bibfield  {author} {\bibinfo {author} {\bibfnamefont {S.~K.}\ \bibnamefont
  {Das}}, \bibinfo {author} {\bibfnamefont {N.}~\bibnamefont {Putra}}, \ and\
  \bibinfo {author} {\bibfnamefont {W.}~\bibnamefont {Roetzel}},\ }\href@noop
  {} {\bibfield  {journal} {\bibinfo  {journal} {International journal of heat
  and mass transfer}\ }\textbf {\bibinfo {volume} {46}},\ \bibinfo {pages}
  {851} (\bibinfo {year} {2003})}\BibitemShut {NoStop}%
\bibitem [{\citenamefont {Wen}\ and\ \citenamefont
  {Ding}(2005)}]{wen2005experimental}%
  \BibitemOpen
  \bibfield  {author} {\bibinfo {author} {\bibfnamefont {D.}~\bibnamefont
  {Wen}}\ and\ \bibinfo {author} {\bibfnamefont {Y.}~\bibnamefont {Ding}},\
  }\href@noop {} {\bibfield  {journal} {\bibinfo  {journal} {Journal of
  Nanoparticle Research}\ }\textbf {\bibinfo {volume} {7}},\ \bibinfo {pages}
  {265} (\bibinfo {year} {2005})}\BibitemShut {NoStop}%
\bibitem [{\citenamefont {Bang}\ and\ \citenamefont
  {Chang}(2005)}]{bang2005boiling}%
  \BibitemOpen
  \bibfield  {author} {\bibinfo {author} {\bibfnamefont {I.~C.}\ \bibnamefont
  {Bang}}\ and\ \bibinfo {author} {\bibfnamefont {S.~H.}\ \bibnamefont
  {Chang}},\ }\href@noop {} {\bibfield  {journal} {\bibinfo  {journal}
  {International Journal of Heat and Mass Transfer}\ }\textbf {\bibinfo
  {volume} {48}},\ \bibinfo {pages} {2407} (\bibinfo {year}
  {2005})}\BibitemShut {NoStop}%
\bibitem [{\citenamefont {Chopkar}\ \emph {et~al.}(2008)\citenamefont
  {Chopkar}, \citenamefont {Das}, \citenamefont {Manna},\ and\ \citenamefont
  {Das}}]{chopkar2008pool}%
  \BibitemOpen
  \bibfield  {author} {\bibinfo {author} {\bibfnamefont {M.}~\bibnamefont
  {Chopkar}}, \bibinfo {author} {\bibfnamefont {A.}~\bibnamefont {Das}},
  \bibinfo {author} {\bibfnamefont {I.}~\bibnamefont {Manna}}, \ and\ \bibinfo
  {author} {\bibfnamefont {P.}~\bibnamefont {Das}},\ }\href@noop {} {\bibfield
  {journal} {\bibinfo  {journal} {Heat and Mass Transfer}\ }\textbf {\bibinfo
  {volume} {44}},\ \bibinfo {pages} {999} (\bibinfo {year} {2008})}\BibitemShut
  {NoStop}%
\bibitem [{\citenamefont {Li}\ \emph {et~al.}(2017)\citenamefont {Li},
  \citenamefont {Gonzalez-Avila}, \citenamefont {Nguyen},\ and\ \citenamefont
  {Ohl}}]{li2017oscillate}%
  \BibitemOpen
  \bibfield  {author} {\bibinfo {author} {\bibfnamefont {F.}~\bibnamefont
  {Li}}, \bibinfo {author} {\bibfnamefont {S.~R.}\ \bibnamefont
  {Gonzalez-Avila}}, \bibinfo {author} {\bibfnamefont {D.~M.}\ \bibnamefont
  {Nguyen}}, \ and\ \bibinfo {author} {\bibfnamefont {C.-D.}\ \bibnamefont
  {Ohl}},\ }\href@noop {} {\bibfield  {journal} {\bibinfo  {journal} {Physical
  Review Fluids}\ }\textbf {\bibinfo {volume} {2}},\ \bibinfo {pages} {014007}
  (\bibinfo {year} {2017})}\BibitemShut {NoStop}%
\bibitem [{\citenamefont {Chen}\ and\ \citenamefont
  {Chung}(2002)}]{chen2002coalescence}%
  \BibitemOpen
  \bibfield  {author} {\bibinfo {author} {\bibfnamefont {T.}~\bibnamefont
  {Chen}}\ and\ \bibinfo {author} {\bibfnamefont {J.}~\bibnamefont {Chung}},\
  }\href@noop {} {\bibfield  {journal} {\bibinfo  {journal} {International
  Journal of Heat and Mass Transfer}\ }\textbf {\bibinfo {volume} {45}},\
  \bibinfo {pages} {2329} (\bibinfo {year} {2002})}\BibitemShut {NoStop}%
\bibitem [{\citenamefont {Deng}\ \emph {et~al.}(2003)\citenamefont {Deng},
  \citenamefont {Lee},\ and\ \citenamefont {Cheng}}]{deng2003design}%
  \BibitemOpen
  \bibfield  {author} {\bibinfo {author} {\bibfnamefont {P.}~\bibnamefont
  {Deng}}, \bibinfo {author} {\bibfnamefont {Y.-K.}\ \bibnamefont {Lee}}, \
  and\ \bibinfo {author} {\bibfnamefont {P.}~\bibnamefont {Cheng}},\ }in\
  \href@noop {} {\emph {\bibinfo {booktitle} {Transducers, Solid-State Sensors,
  Actuators and Microsystems, 12th International Conference on, 2003}}},\
  Vol.~\bibinfo {volume} {1}\ (\bibinfo {organization} {IEEE},\ \bibinfo {year}
  {2003})\ pp.\ \bibinfo {pages} {647--650}\BibitemShut {NoStop}%
\bibitem [{\citenamefont {Deng}\ \emph {et~al.}(2006)\citenamefont {Deng},
  \citenamefont {Lee},\ and\ \citenamefont {Cheng}}]{deng2006experimental}%
  \BibitemOpen
  \bibfield  {author} {\bibinfo {author} {\bibfnamefont {P.}~\bibnamefont
  {Deng}}, \bibinfo {author} {\bibfnamefont {Y.-K.}\ \bibnamefont {Lee}}, \
  and\ \bibinfo {author} {\bibfnamefont {P.}~\bibnamefont {Cheng}},\
  }\href@noop {} {\bibfield  {journal} {\bibinfo  {journal} {International
  Journal of Heat and Mass Transfer}\ }\textbf {\bibinfo {volume} {49}},\
  \bibinfo {pages} {2535} (\bibinfo {year} {2006})}\BibitemShut {NoStop}%
\bibitem [{\citenamefont {Chen}\ \emph {et~al.}(2006)\citenamefont {Chen},
  \citenamefont {Klausner}, \citenamefont {Garimella},\ and\ \citenamefont
  {Chung}}]{chen2006subcooled}%
  \BibitemOpen
  \bibfield  {author} {\bibinfo {author} {\bibfnamefont {T.}~\bibnamefont
  {Chen}}, \bibinfo {author} {\bibfnamefont {J.~F.}\ \bibnamefont {Klausner}},
  \bibinfo {author} {\bibfnamefont {S.~V.}\ \bibnamefont {Garimella}}, \ and\
  \bibinfo {author} {\bibfnamefont {J.~N.}\ \bibnamefont {Chung}},\ }\href@noop
  {} {\bibfield  {journal} {\bibinfo  {journal} {International journal of heat
  and mass transfer}\ }\textbf {\bibinfo {volume} {49}},\ \bibinfo {pages}
  {4399} (\bibinfo {year} {2006})}\BibitemShut {NoStop}%
\bibitem [{\citenamefont {Xu}\ and\ \citenamefont
  {Zhang}(2008)}]{xu2008effect}%
  \BibitemOpen
  \bibfield  {author} {\bibinfo {author} {\bibfnamefont {J.}~\bibnamefont
  {Xu}}\ and\ \bibinfo {author} {\bibfnamefont {W.}~\bibnamefont {Zhang}},\
  }\href@noop {} {\bibfield  {journal} {\bibinfo  {journal} {International
  Journal of Heat and Mass Transfer}\ }\textbf {\bibinfo {volume} {51}},\
  \bibinfo {pages} {389} (\bibinfo {year} {2008})}\BibitemShut {NoStop}%
\bibitem [{\citenamefont {Romera-Guereca}\ \emph {et~al.}(2008)\citenamefont
  {Romera-Guereca}, \citenamefont {Choi},\ and\ \citenamefont
  {Poulikakos}}]{romera2008explosive}%
  \BibitemOpen
  \bibfield  {author} {\bibinfo {author} {\bibfnamefont {G.}~\bibnamefont
  {Romera-Guereca}}, \bibinfo {author} {\bibfnamefont {T.}~\bibnamefont
  {Choi}}, \ and\ \bibinfo {author} {\bibfnamefont {D.}~\bibnamefont
  {Poulikakos}},\ }\href@noop {} {\bibfield  {journal} {\bibinfo  {journal}
  {International Journal of Heat and Mass Transfer}\ }\textbf {\bibinfo
  {volume} {51}},\ \bibinfo {pages} {4427} (\bibinfo {year}
  {2008})}\BibitemShut {NoStop}%
\bibitem [{Note1()}]{Note1}%
  \BibitemOpen
  \bibinfo {note} {See supplementary materials Fig.S1 and Fig.S2 for detailed
  fabrication process and thermal simulation of the electrical
  heater}\BibitemShut {NoStop}%
\bibitem [{\citenamefont {Baumeister}\ and\ \citenamefont
  {Simon}(1973)}]{baumeister1973leidenfrost}%
  \BibitemOpen
  \bibfield  {author} {\bibinfo {author} {\bibfnamefont {K.}~\bibnamefont
  {Baumeister}}\ and\ \bibinfo {author} {\bibfnamefont {F.}~\bibnamefont
  {Simon}},\ }\href@noop {} {\bibfield  {journal} {\bibinfo  {journal}
  {(American Society of Mechanical Engineers, 1973.) ASME, Transactions, Series
  C- Journal of Heat Transfer,}\ }\textbf {\bibinfo {volume} {95}},\ \bibinfo
  {pages} {166} (\bibinfo {year} {1973})}\BibitemShut {NoStop}%
\bibitem [{Note2()}]{Note2}%
  \BibitemOpen
  \bibinfo {note} {See supplementary materials Video 2 for the recordings of
  this process}\BibitemShut {NoStop}%
\bibitem [{Note3()}]{Note3}%
  \BibitemOpen
  \bibinfo {note} {See Supplementary Materials section II.C for further
  discussion on the unstable nature of this regime.}\BibitemShut {Stop}%
\bibitem [{Note4()}]{Note4}%
  \BibitemOpen
  \bibinfo {note} {See supplementary materials part II.D for detailed
  comparisons of two heating approaches.}\BibitemShut {Stop}%
\bibitem [{Note5()}]{Note5}%
  \BibitemOpen
  \bibinfo {note} {See Supplementary Materials Fig. S7 for results with 10$\mu
  $m $\times $ 10$\mu $m heater.}\BibitemShut {Stop}%
\end{thebibliography}%

\end{document}